\documentclass[aps,prl,twocolumn,english,superscriptaddress,10pt]{revtex4-1}
\usepackage{verbatim}
\usepackage{amsmath}
\usepackage{amssymb}
\usepackage{graphicx}
\usepackage{braket}
\usepackage{xcolor,subfigure,upgreek,bbold}
\usepackage[colorlinks,linkcolor=blue,citecolor=blue,urlcolor=blue,breaklinks=true]{hyperref}
\usepackage{verbatim}
\usepackage{esint}

\graphicspath{{./figs/}{./}}

\begin{document}

\title{Spin-models, dynamics and criticality with atoms in tilted optical superlattices}

\author{Anton S. Buyskikh}
\email{anton.buyskikh@strath.ac.uk}
\affiliation{Department of Physics and SUPA, University of Strathclyde, Glasgow G4 0NG, UK}

\author{Luca Tagliacozzo}
\affiliation{Department of Physics and SUPA, University of Strathclyde, Glasgow G4 0NG, UK}
\affiliation{Departament de F\'{\i}sica Qu\`antica i Astrof\'{\i}sica and Institut de Ci\`encies del Cosmos (ICCUB), Universitat de Barcelona,  Mart\'{\i} i Franqu\`es 1, 08028 Barcelona, Catalonia, Spain}

\author{Dirk Schuricht}
\affiliation{Institute for Theoretical Physics, Center for Extreme Matter and Emergent Phenomena, Utrecht University, Princetonplein 5, 3584 CE Utrecht, The Netherlands}

\author{Chris A. Hooley}
\affiliation{SUPA, School of Physics and Astronomy, University of St Andrews, North Haugh, St Andrews KY16 9SS, UK}

\author{David Pekker}
\affiliation{Department of Physics and Astronomy, University of Pittsburgh, Pittsburgh, PA 15260, USA}

\author{Andrew J. Daley}
\affiliation{Department of Physics and SUPA, University of Strathclyde, Glasgow G4 0NG, UK}

\date{\today}
\pacs{...}

\begin{abstract}
We show that atoms in tilted optical superlattices provide a platform for exploring coupled spin chains of forms that are not present in other systems. In particular, using a period-2 superlattice in 1D, we show that coupled Ising spin chains with XZ and ZZ spin coupling terms can be engineered. We use optimized tensor network techniques to explore the criticality and non-equilibrium dynamics in these models, finding a tricritical Ising point in regimes that are accessible in current experiments. These setups are ideal for studying low-entropy physics, as initial entropy is ``frozen-out'' in realizing the spin models, and provide an example of the complex critical behaviour that can arise from interaction-projected models.
\end{abstract}

\maketitle

Systems of ultracold atoms are regularly discussed and utilized as quantum simulators to engineer models of interest from other many-body physical systems that are computationally complex \cite{Bloch2012,Lewenstein2012,Jaksch2005,Jaksch1998}. However, the unique properties of these systems can also motivate the study of genuinely novel many-body physics --- especially including a wide variety of phenomena in out-of-equilibrium dynamics \cite{Polkovnikov2011,Daley2014b,Muller2012}. Here, we explore spin models that arise naturally for ultracold atoms loaded in an optical superlattice potential together with a gradient potential. The large energy scales provided by the gradient produce a dynamical constraint, similar to projected models arising from large-scale interactions, which have recently been discussed in arrays of Rydberg atoms \cite{Choi2018,Ho2019,Samajdar2018,Verresen2019} and topological wires \cite{Surace2019}. We show how these models, which can be rewritten as coupled spin chains with unusual couplings, exhibit complex critical behaviour that can be explored both in and out of equilibrium.

\begin{figure}[tb]
\begin{centering}
\includegraphics[width=1\linewidth]{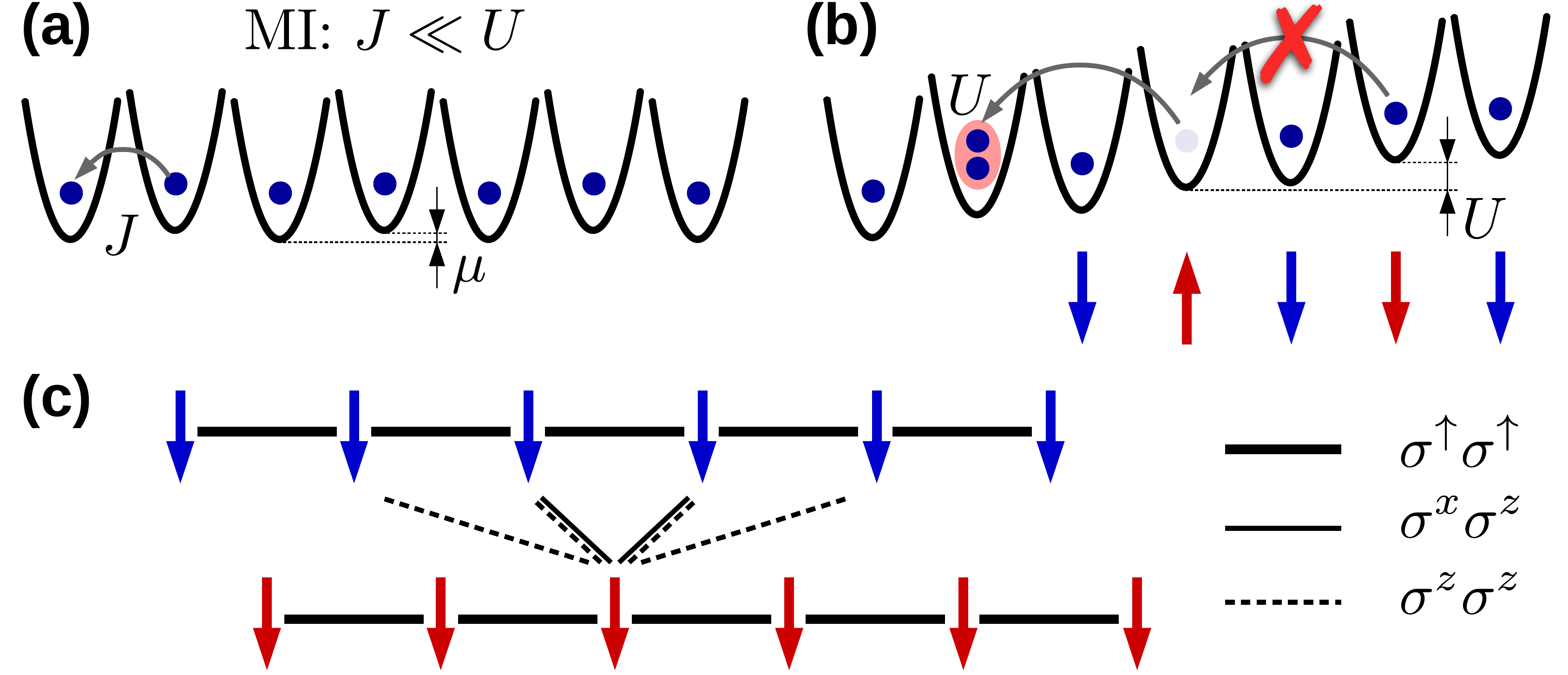}
\par\end{centering}
\caption{\label{fig:diagram}
Spin model example. (a) We consider states that are resonantly coupled to a unit filled MI phase of bosons on a tilted superlattice with period 2 and an energy offset $\mu$.
(b) For atoms resonantly tunnelling over two sites, $E \approx U/2$, the superlattice provides a closed spin-1/2 model, where a spin is labelled $\ket{\uparrow}$ if the particle has tunnelled by two sites, and $\ket{\downarrow}$ if it has not tunnelled. There is a strong constraint preventing two consecutive spin-up states in any sublattice, as this state is not resonantly coupled. 
(c) The resulting effective spin model \eqref{eq:Heff_U2} is represented as two separate Ising spin chains, one for each sublattice. This setup provides unusual couplings between these spin chains: chains of odd (blue) and even (red) spins coupled with each other via $\sigma^x\sigma^z$ and $\sigma^z\sigma^z$ interactions (shown only for one spin).}
\end{figure}

Previous experiments have shown long-lived coherent dynamics of strongly interacting atoms confined to move along one dimension of a tilted optical lattice \cite{Greiner2002,Simon2011,Meinert2013,Meinert2014}. The origin of these coherences was investigated in \cite{Sachdev2002}, which considered all states resonantly coupled to that with a single atom in every lattice site, then when the energy difference along the potential gradient between neighboring lattice sites, $E$, is equal to the on-site interaction energy shift $U$, the system was mapped onto an effective skew-field Ising model. Our work is strongly motivated by experiments observing resonant dynamics for $E=U/n$ with $n=1,2,3,\ldots,6$ \cite{Meinert2014}. Here we show that by adding a superlattice potential, it is possible to make use of these resonant dynamics to represent an unusual form of coupled Ising spin chains. Using symmetry arguments and computational techniques with Matrix Product Operators we identify critical behavior, which we characterize as a tricritical point in the phase diagram, and analyze characteristic quench dynamics that would be directly observable in ongoing experiments.

\emph{Model} -- Atoms confined to the lowest band of a sufficiently deep 1D optical lattice are quantitatively described by the modified Bose-Hubbard model ($\hbar\equiv 1$)
\begin{equation}
H=-J\sum_{\left\langle i,j \right\rangle}b_i^\dagger b_j+\frac{U}{2}\sum_i n_i (n_i-1)-\sum_i V_i n_i,\label{eq:H_BH}
\end{equation}
where $J$ is the tunnelling amplitude, $\braket{\cdot,\cdot}$ implies summation over nearest neighbors, and $V_{i}=E\cdot i+(-1)^i \cdot \mu/2$ is the linear energy shift with the superlattice energy offset $\mu$, crucial for the $n=2$ spin model derivation. Further details of the following derivations are presented in \cite{Buyskikh2018a}.

By loading atoms into a deep lattice with $U \gg J$, we begin from a unit filled Mott Insulator (MI) state with atoms localized at individual lattice sites. When the tilt is tuned to the vicinity of a resonance $E=U/n$, $n=1,2,3,\ldots$ long-range resonantly enhanced tunnelings of $n^\mathrm{th}$ order become possible on the experimental time scales \cite{Simon2011, Meinert2013, Meinert2014}.The original study for $n=1$ \cite{Sachdev2002} proposed a mapping between bosonic spatial degrees of freedom and effective spins. Each boson that stays on its initial site of the MI state is mapped with $\ket{\downarrow}$ and each boson that resonantly hops to the first neighboring site as $\ket{\uparrow}$.

In the regime $\left|U-E\right|,J\ll E,U$ and $\mu=0$ the behavior at relevant time scales is mapped to the effective model
\begin{equation}
H_{U}=\sum_i\big[-\sqrt{2}\sigma_i^x+\tilde\lambda\sigma_i^\uparrow+W\sigma_i^\uparrow\sigma_{i+1}^\uparrow\big],\label{eq:Heff_U1}
\end{equation}
where $\tilde \lambda=(U-E)/J$ denotes the deviation from the resonance, $\sigma^{\uparrow}=(\sigma^z+1)/2$ is a projector on the spin-up state, and $W\to+\infty$ is the constraint preventing two consecutive spin-ups, as this configuration would not be resonantly coupled and has occupation suppressed $\propto(J/U)^2$.
In later works non-equilibrium dynamical properties \cite{Kolodrubetz2012} and high order corrections \cite{Munoz-Arias2016} were investigated within this model. The model \eqref{eq:Heff_U1} is referred as the antiferromagnetic Ising chain in a skew field (AFISF), which exhibits two phases: an ordered AFM phase and paramagnetic (PM) phase with spins aligned with the skew field \cite{Ovchinnikov2003}. These two phases are separated by a second order phase transition line on the plane of transverse and parallel fields, reflecting breaking of $\mathcal{Z}_2$ symmetry.

\emph{Spin model for $n=2$} -- For $n>1$ we naturally consider higher-order tunnelling processes in the small parameter $J/E \sim J/(U/n)$, which have been experimentally observed in \cite{Meinert2014}.
It is necessary to introduce the superlattice potential for $n=2$ to obtain an effective spin model, as otherwise second order tunnelling processes give rise to atom configurations that are not describable by any spin-half model. 

Using this mapping in the regime $|U/2-E|,J\ll\mu\ll E,U$ we obtain an effective Hamiltonian of the following form, as depicted in Fig.~\ref{fig:diagram}(c),
\begin{gather}
H_{U/2}=\sum_i \Big[-\sqrt{2}\sigma_i^x+\lambda\sigma_i^\uparrow+W\sigma_i^\uparrow\sigma_{i+2}^\uparrow+\frac{8-56\sigma_i^\uparrow}{15}\nonumber \\
-2\sqrt{2}(\sigma_i^x\sigma_{i+1}^z+\sigma_i^z\sigma_{i+1}^x)-\frac{8}{5}(\sigma_i^z\sigma_{i+1}^z+\sigma_i^z\sigma_{i+3}^z)\Big].\label{eq:Heff_U2}
\end{gather}
Here, $\lambda=\frac{U/2-E}{J^{2}/(U/2)}$ denotes the deviation from the resonance. For the spin model, this has the same form as $\tilde \lambda$ for the $n=1$ case, but involves the second order tunnelling amplitude and detuning from resonance of the current case. 
The constraint $W\to+\infty$ forbids states with $\ket{\uparrow}_i\ket{\uparrow}_{i+2}$, i.e., two neighboring atoms within a sublattice cannot both tunnel, analogously to the $n=1$ case (with maximum occupation of the states we project out scaling as $\propto(J/U)^4$). The first three terms alone are identical to \eqref{eq:Heff_U1} for each sublattice, providing two Ising subchains. The fourth term just shifts the entire energy spectrum along the energy and detuning $\lambda$ axes due to interactions between even and odd spins. 

The remaining terms result from coupling between these subchains in second order perturbation theory, and can be understood intuitively as follows. Firstly, the terms proportional to $\sigma^x\sigma^z$ arise because as atoms tunnel within each sublattice (represented by $\sigma^x$) the denominator of the resulting amplitude in second order perturbation theory will change sign depending on whether an atom is present on the intermediate site from the other sublattice (giving a $\sigma^z$ on the neighboring site). The $\sigma^z\sigma^z$ terms arise due to virtual energy shifts in second-order perturbation theory that are dependent on the occupation number in given sites. These can be separated by up to three sites in the spin model, because an atom can tunnel over two sites and encounter a shift depending on whether an atom is present on the neighboring (third) site.

Despite the relatively simple couplings, we are not aware of any other physical realization of this type of spin model. We analyzed its properties through analytical arguments and numerical calculations based on Matrix Product State and Operators (MPS/MPO) techniques \cite{Schollwoeck2011}, with all states converged in a matrix product bond dimension $D$. Throughout our calculations we implement the constraint in \eqref{eq:Heff_U1} and \eqref{eq:Heff_U2} exactly with matrix product projection operators, as detailed in \cite{Buyskikh2018a}. Finite temperature and out of equilibrium calculations were performed using a combination of Time-Dependent Variational Principle (TDVP) methods \cite{Haegeman2011,Koffel2012,Haegeman2013,Haegeman2016} and purification techniques \cite{Verstraete2004,Cuevas2013}. 

\emph{Phase diagram and critical behavior} -- The essential points of the phase diagram become clear if we first discuss the extreme cases.
As in the $n=1$ case, for $\lambda\to+\infty$ the ground state of the system is non-degenerate PM state $\prod_i \ket{\downarrow}_i$ where spins are aligned with the strong external field $\lambda$, in the bosonic language the energy is minimized if all bosons stay on their initial sites of the parent MI state.
In the case of $\lambda\to-\infty$ spins in the ground state are ordered in the AFM fashion $\ket{(\downarrow\downarrow\uparrow\uparrow)}$, where odd and even spins are Neel ordered due to the constraint.

\begin{figure}[tb]
\includegraphics[width=1\linewidth]{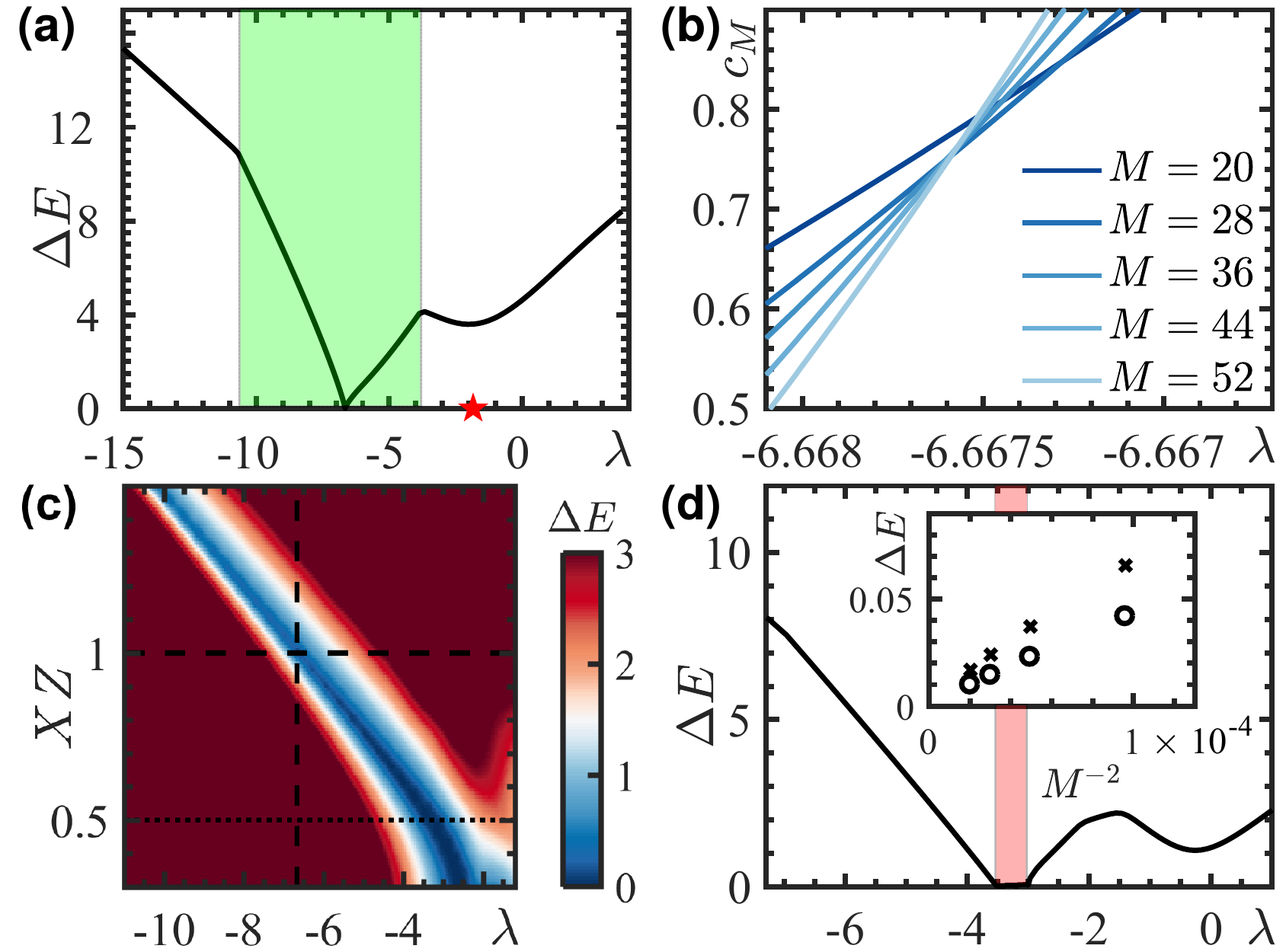}
\caption{
\label{fig:en_gap}
(a)
Energy gap $\Delta E$ vs. $\lambda$ from Eq.~\eqref{eq:Heff_U2}.
We choose OBC and $M=142$ spins such that the system always has a non-degenerate ground state in AFM and PM phases.
This energy gap scaling example shows that the nature of the lowest excitations changes near the QCP.
Outside of the shaded green area the gap behavior is described by elementary excitations identical for the regime $E=U$ \cite{Sachdev2002}.
Inside, the gap scales as $\sim2\lambda$ as bound excitation pairs have lower energy than elementary excitations.
The red star denotes the QCP in the regime $E=U$ \cite{Sachdev2002} for comparison.
(b)
The central charge sequence \eqref{eq:central_charge} near the critical point of the spin model \eqref{eq:Heff_U2} with PBC.
The intersection points $\lambda_{\times}(M_{\times})$ converge to $\lambda_\mathrm{crit}\approx-6.6676(1)$ (see the main text).
The maximal discrepancy in $\lambda_{\times}$ between MPS bond dimensions $D=384$ and 512 is $2\cdot10^{-5}$.
(c)
Energy gap $\Delta E$ vs $\lambda$ and the relative strength of $\sigma^x \sigma^z$ interactions of Eq.~\eqref{eq:Heff_U2} for OBC and $M=102$ spins.
The position of the tricritical point is marked as the intersection of dashed lines.
(d)
The same as (a) but with $XZ=1/2$ (dotted line in (c)).
The red shaded area highlights a new phase area. The inset shows $\Delta E$ scaling for two distinctly different $\lambda$.
Calculations were performed using DMRG and eMPS methods \cite{Wall2012}, numerical convergence was achieved for bond dimensions not larger than $D=512$.
The restricted spin configurations were excluded from the calculations via an MPO projector (see Appendix of \cite{Buyskikh2018a}).}
\end{figure}

The lowest elementary excitations are single spin flips $\ket{j}=\ket{\uparrow}_j \prod_{i\neq j}\ket{\downarrow}_i$ for the PM and domain walls for the AFM, with the energy gap over the ground state scaling as $\sim\lambda$ (or $\tilde \lambda$), see \cite{Buyskikh2018a} for the microscopic picture details. Coupling between the chains, however, lead to bound excitations on the two subchains, and this is enough to suppress the original phase transition of elementary excitations at $\tilde\lambda=-1.853$ and create a new phase transition point, which can be seen in Fig.~\ref{fig:en_gap}(a).

We have found that the low-energy physics of \eqref{eq:Heff_U2} is strongly influenced by interactions between the bound excitations. It is  very sensitive to the precise location of the critical point $\lambda_\mathrm{crit}$ rendering the identification of the universality class of the transition very challenging \cite{Buyskikh2018a}.
We thus use a technique that, similarly to Binder cumulants \cite{Binder1981}, makes it possible to locate the critical point without any previous knowledge of the critical exponents beside assuming that the theory becomes conformally invariant.
In this way we can use a prediction from conformal field theory (CFT) regarding the scaling of the entanglement entropy. For blocks of contiguous $M$ spins embedded in a chain with periodic boundary conditions, we expect \cite{Callan1994,Osborne2002,Vidal2003,Calabrese2004}
\begin{equation}
S_{\mathrm{vN}}(M) = \frac{c}{3} \log_2 M + A + \dots,
\label{eq:svn}
\end{equation}
where $c$ is the central charge of the corresponding CFT \cite{Affleck1986,Blote1986,Cardy2010a}, $A$ is a non-universal constant and the dots suggest the possible presence of further sub-leading corrections \cite{Laflorencie2006,Calabrese2010,Calabrese2010a,Cardy2010,Xavier2012}.

We can then extract the central charge at the critical point by plotting
\begin{equation}
c_M=3\frac{S_\mathrm{vN}(M)-S_\mathrm{vN}(M_{\max})}{\log_2(M/M_{\max})},\label{eq:central_charge}
\end{equation}
for system sizes $M\leq 60$ vs. $\lambda$, Fig.~\ref{fig:en_gap}(b) \footnote{This method was initially proposed in the supplementary material of \cite{Koffel2012}.}.
All system sizes $M$ considered here are chosen to be divisible by 4, the periodicity of the AFM phase.
This ensures that the ground states have the same degeneracy and hence corrections related to different degeneracies can be neglected.

At the Quantum Critical Point (QCP) the sequence $c_M$ (i) becomes independent of $M$ and (ii) matches the value of the central charge $c$.
In order to locate the critical point $\lambda_\mathrm{crit}$ we perform the following analysis.
For each pair of lines $c_{M}$ and $c_{M+4}$ we find the intersection point $(M_\times,\lambda_\times)$, where $M_\times=M+2$.
This sequence $\lambda_\times(M_\times)$ approaches a stationary value that gives the best estimate for the critical point $\lambda_\mathrm{crit}\approx-6.6676(1)$ (Fig.~\ref{fig:en_gap}(b)).
The maximal discrepancy in $\lambda_{\times}$ between MPS bond dimensions $D=384$ and 512 is $2 \cdot 10^{-5}$, which does not significantly contribute to this error estimate.
From Fig.~\ref{fig:en_gap}(b) we estimate the central charge $c\approx0.75$, however the presence of sub-leading correction in $S_\mathrm{vN}$ smears out the crossing points.

Since the value of the central charge is below unity, its value should match at least one tabulated value of a unitary representation of the Virasoro algebra \cite{Belavin1984,Henkel1999,Francesco1997,Cardy2008}.
The two closest values of the central charge that correspond to a minimal model are $c=7/10$ (tricritical Ising point) and $c=4/5$ (3-state Potts model). We observe an extreme sensitivity of the central charge value to the critical point location. From Fig.~\ref{fig:en_gap}(b) we see that the dependence of $c_M$ on $\lambda$ becomes extremely steep at the critical point as $M$ increases. For instance, the estimated error of the critical point is $~10^{-4}$, which leads to the central charge uncertainty $c_M\approx0.7-0.8$.
This phenomenon of high sensitivity is absent at a standard critical point (such as in the 3-state Potts model), and has already been observed in the context of the tricritical Ising point \cite{Xavier2012}. Furthermore, at a tricritical point three distinct phases merge and there are two relevant symmetry breaking fields.
We thus expect that if the system is tricritical our $\lambda$ is actually tuning two relevant parameters.

In order to understand the existence of a tricritical point in this model, we investigated the energy gap while adjusting the $XZ$ parameter (Fig.~\ref{fig:en_gap}(c,d)) representing the \textit{relative} strength of $\sigma^x \sigma^z$ interactions, i.e. the regime with $XZ=1$ is identical to the Hamiltonian~\eqref{eq:Heff_U2}.
There we see a region appears where for a range of $\lambda$ values the energy gap scales to zero with increasing system size.
This suggests the existence of a third phase in the model that the ground state becomes either highly degenerate or a gapless phase of bound excitations.
Another possibility is that with the weakened $\sigma^x\sigma^z$ interactions a new floating incommensurate phase \cite{Fendley2004,Chepiga2019} emerges, which is caused by the competition between the constraint and couplings. 
This would be an interesting direction for future investigations.

\begin{figure}[tb]
\begin{centering}
\includegraphics[width=1\linewidth]{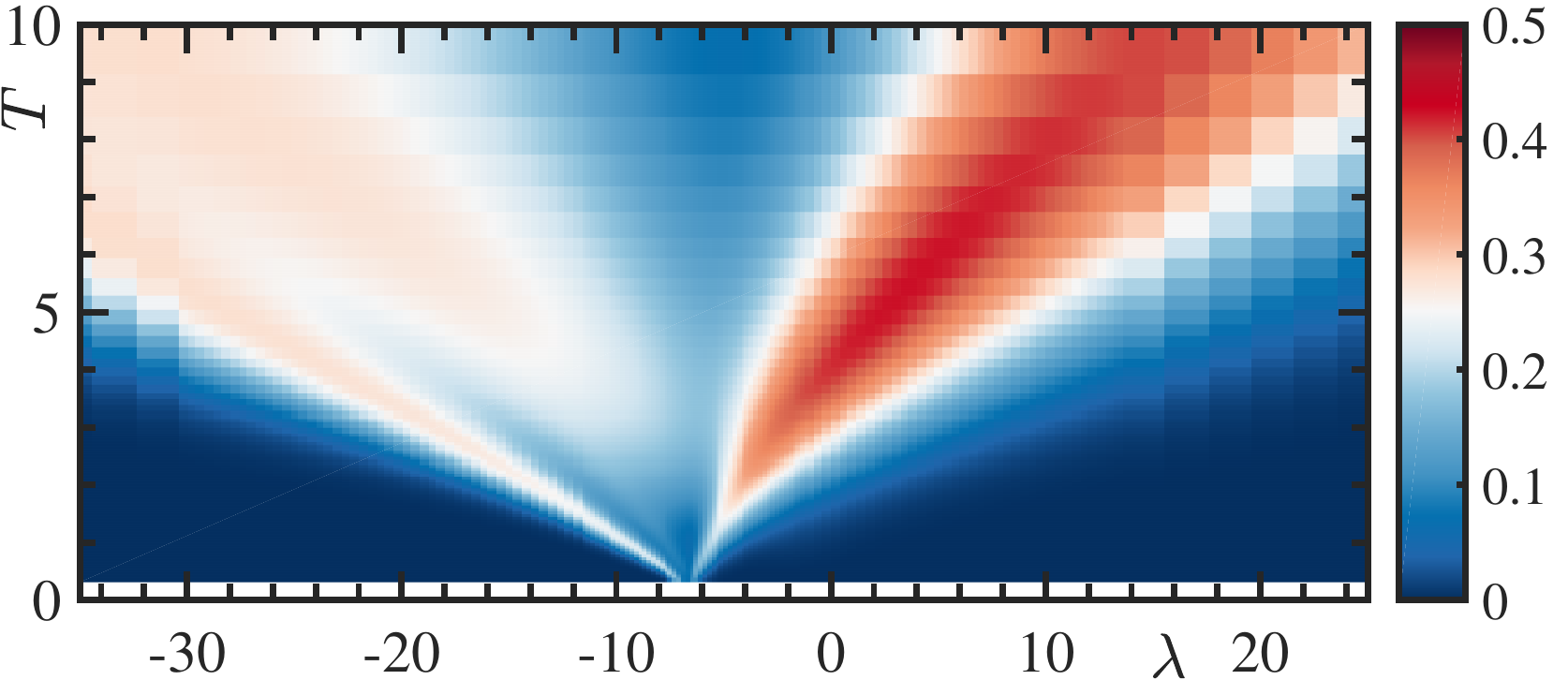}
\par\end{centering}
\caption{\label{fig:Cv_U2}
Specific heat capacity per spin \eqref{eq:Cv} for the system \eqref{eq:Heff_U2} of $M=102$ spins with OBC.
Two types of excitation observed in Fig.~\ref{fig:en_gap}(a) both approach the ground state energy.
Coupled excitations converge to $\lambda_\mathrm{crit}$, which can be seen here as converging specific heat branches.
Whereas elementary excitations always have a finite gap and their branches vanish at finite temperatures.
Proximity of the transitions cause two of four branches to overlap.
The results are obtained via imaginary time evolution of the infinite-$T$ density matrix via TDVP method with MPS with bond dimension $D=128$.}
\end{figure}

\emph{Specific heat} -- To connect more closely with how this might be observed in experiments, using MPO techniques to compute the finite-temperature behavior of the system, we determine the specific heat capacity per spin
\begin{equation}
c(T,\lambda)=\braket{\Delta H_{U/2}^2(\lambda)}_T/T^2 M.\label{eq:Cv}
\end{equation}
Here $\braket{\Delta H_{U/2}^2(\lambda)}_T$ is the total energy variance of the Hamiltonian \eqref{eq:Heff_U2} at temperature $T$, and shows spectral lines accessible experimentally, e.g., in spectroscopy via modulation of external fields.
As shown in Fig.~\ref{fig:Cv_U2}, this is significantly more complicated than we would expect for an Ising critical point.
The dependence of the specific heat can be understood with a help of the energy gap in Fig.~\ref{fig:en_gap}(a), where we see two transition points: the quantum phase transition of bounded excitations, and avoided transition due to elementary excitations (used to be the phase transition of \eqref{eq:Heff_U1}).
Each transition has two branches of specific heat maxima diverging from it due to thermal fluctuations. We expect that Fig.~\ref{fig:Cv_U2} actually contains four branches, but for positive $\lambda$, two partly overlap and cannot be distinguished. 

\begin{figure}[tb]
\begin{centering}
\includegraphics[width=1\linewidth]{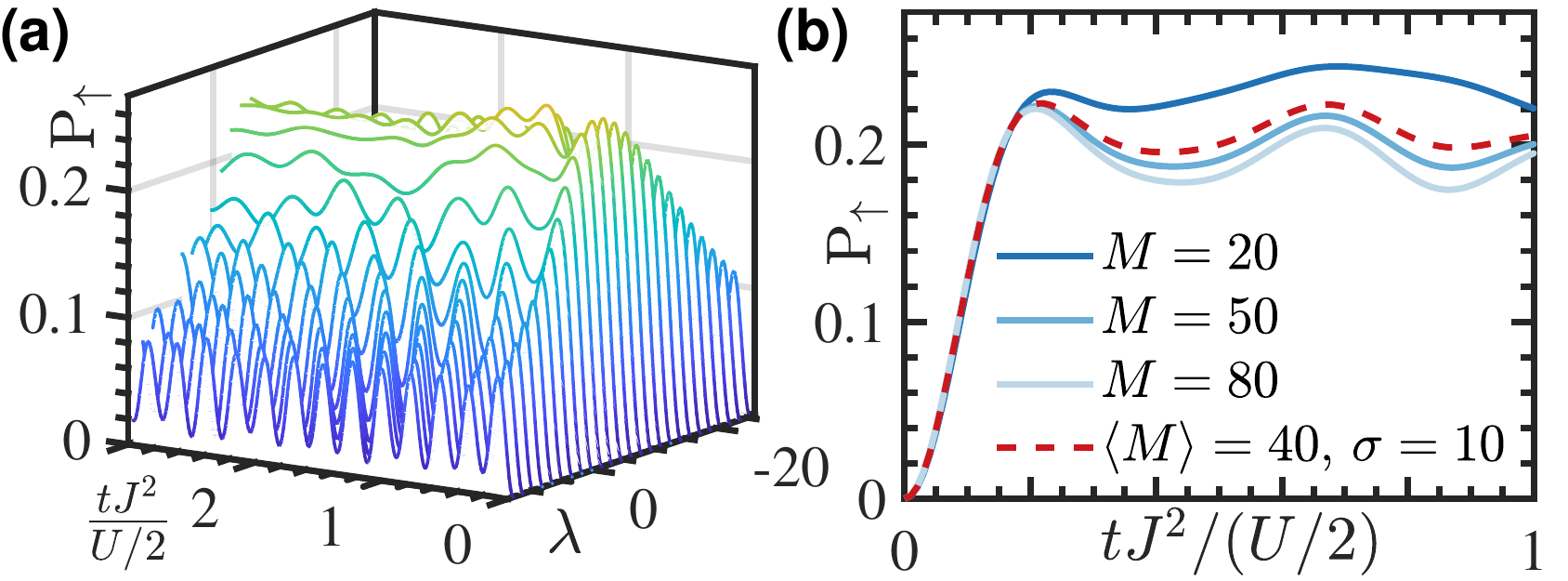}
\par\end{centering}
\caption{\label{fig:quenches}
Evolution of the average number of up-spins $\mathrm{P}_\uparrow=\sum_i \braket{\sigma_i^\uparrow}/M$, which maps onto double occupied sites, after instantaneous global quenches from the unit filled MI state to the regime $E=U/2$ \eqref{eq:Heff_U2}.
(a) A collection of quenches with different detuning $\lambda$ for $M=50$ spins and OBC.
(b) The time evolution at $\lambda=0$ for spin chains of length $M$ (solid lines) and for an ensemble of spin chains, the lengths of which are normally distributed with the mean $\braket{M}=40$ and standard deviation $\sigma=10$ (dashed line), simulating the experiment where many 1D lattices are probed in parallel.
Calculations were performed using TDVP method with MPS with bond dimension $D=128$.}
\end{figure}

\emph{Quench dynamics} -- In experiments, the most straight-forward way to probe the transition will be via quench dynamics, including finite-speed analysis of quenches \cite{DeGrandi2011, Kolodrubetz2012}, and generation of excitations (which could also be probed in quantum gas microscopes). In Fig.~\ref{fig:quenches} we present the dynamics of \eqref{eq:Heff_U2} after sudden quenches, which is mapped to a quench from a unit filled MI with $E=0$ to a point near the $E=U/2$ resonance. After a sudden quench we observe resonant behavior in the number of up-spins, in analogy with dynamics at the $E=U$ resonance \cite{Rubbo2011,Meinert2013}. We also see clear overdamping of density oscillations, in contrast to the $E=U$ resonance \cite{Meinert2014}. This occurs primarily because the $\sigma^z \sigma^z$ coupling between subchains results in a large number of frequency components in the dynamics, separated by small values of the order of the $\sigma^x$ terms in the individual chains. The quench dynamics results in the spin system in saturation timescale and mean saturation number of doublons closely resemble the experimental results in the regime $E=U/2$ without a superlattice \cite{Meinert2014}. This indicates that the underlying physics may be related, even though the models are different, and these connections could be more closely explored in future experiments.

\emph{Summary and outlook} -- We have shown that atoms in tilted superlattices provide a new opportunity to realize coupled spin models based on the atom number. This work opens avenues for broader investigations of projected models, both for itinerant atoms, with extensions to more coupled chains with longer period superlattices, 2D \cite{Pielawa2012,Kolovsky2016}, or tunnelling of spinful fermions and bosons; and also for arrays of Rydberg atoms and polar molecules, allowing the engineering of distance-dependent spin-spin interactions. Our work provides an example of the complex critical behaviour that can emerge from projected models, and specifically provides a physical realization of tricritical Ising behaviour that has been of significant recent interest in conformal field theory \cite{bootstrapcollaboration,Poland2019,Atanasov2018,OBrien2018} and for interacting Majorana fermions \cite{Zhu2016}. Another aspect of this work is in connection with commensurate-incommensurate transitions caused by the presence of chirality due to domain walls of different types \cite{Huse1982} between four different vacuum states in the symmetry broken phase. Its experimental realization will provide new insight in the physics of such domain walls, and within quantum gas microscopes, these could be observed directly. The optimized tensor network techniques used here, which are further detailed in \cite{Buyskikh2018a} are also applicable to a wide range of systems, including the simulation of lattice gauge theories \cite{Tagliacozzo2014,Haegeman2015,Rico2014}.

The data for this manuscript is available in open access at Ref.~\cite{data_short}.

\begin{acknowledgements}

We thank Pasquale Calabrese, Florian Meinert, Manfred Mark, Giuseppe Mussardo, Roger Mong, Hanns-Christoph N\"agerl, Subir Sachdev, and Jon Simon for helpful and stimulating discussions. Work at the University of Strathclyde was supported by the EPSRC Programme Grant DesOEQ (EP/P009565/1), by the European Union Horizon 2020 collaborative project QuProCS - Quantum Probes for Complex Systems (grant agreement 641277), and by the EOARD via AFOSR grant number FA2386-14-1-5003. D.S.~acknowledges support of the D-ITP consortium, a program of the Netherlands Organisation for Scientific Research (NWO) that is funded by the Dutch Ministry of Education, Culture, and Science (OCW). D.P.~was supported by the Charles E Kaufman foundation. Results were obtained using the EPSRC funded ARCHIE-WeSt High Performance Computer (EP/K000586/1).

\end{acknowledgements}

\bibliography{references}

\begin{thebibliography}{63}%
\makeatletter
\providecommand \@ifxundefined [1]{%
 \@ifx{#1\undefined}
}%
\providecommand \@ifnum [1]{%
 \ifnum #1\expandafter \@firstoftwo
 \else \expandafter \@secondoftwo
 \fi
}%
\providecommand \@ifx [1]{%
 \ifx #1\expandafter \@firstoftwo
 \else \expandafter \@secondoftwo
 \fi
}%
\providecommand \natexlab [1]{#1}%
\providecommand \enquote  [1]{``#1''}%
\providecommand \bibnamefont  [1]{#1}%
\providecommand \bibfnamefont [1]{#1}%
\providecommand \citenamefont [1]{#1}%
\providecommand \href@noop [0]{\@secondoftwo}%
\providecommand \href [0]{\begingroup \@sanitize@url \@href}%
\providecommand \@href[1]{\@@startlink{#1}\@@href}%
\providecommand \@@href[1]{\endgroup#1\@@endlink}%
\providecommand \@sanitize@url [0]{\catcode `\\12\catcode `\$12\catcode
  `\&12\catcode `\#12\catcode `\^12\catcode `\_12\catcode `\%12\relax}%
\providecommand \@@startlink[1]{}%
\providecommand \@@endlink[0]{}%
\providecommand \url  [0]{\begingroup\@sanitize@url \@url }%
\providecommand \@url [1]{\endgroup\@href {#1}{\urlprefix }}%
\providecommand \urlprefix  [0]{URL }%
\providecommand \Eprint [0]{\href }%
\providecommand \doibase [0]{http://dx.doi.org/}%
\providecommand \selectlanguage [0]{\@gobble}%
\providecommand \bibinfo  [0]{\@secondoftwo}%
\providecommand \bibfield  [0]{\@secondoftwo}%
\providecommand \translation [1]{[#1]}%
\providecommand \BibitemOpen [0]{}%
\providecommand \bibitemStop [0]{}%
\providecommand \bibitemNoStop [0]{.\EOS\space}%
\providecommand \EOS [0]{\spacefactor3000\relax}%
\providecommand \BibitemShut  [1]{\csname bibitem#1\endcsname}%
\let\auto@bib@innerbib\@empty
\bibitem [{\citenamefont {Bloch}\ \emph {et~al.}(2012)\citenamefont {Bloch},
  \citenamefont {Dalibard},\ and\ \citenamefont {Nascimb\`ene}}]{Bloch2012}%
  \BibitemOpen
  \bibfield  {author} {\bibinfo {author} {\bibfnamefont {I.}~\bibnamefont
  {Bloch}}, \bibinfo {author} {\bibfnamefont {J.}~\bibnamefont {Dalibard}}, \
  and\ \bibinfo {author} {\bibfnamefont {S.}~\bibnamefont {Nascimb\`ene}},\
  }\href {\doibase 10.1038/nphys2259} {\bibfield  {journal} {\bibinfo
  {journal} {Nat. Phys.}\ }\textbf {\bibinfo {volume} {8}},\ \bibinfo {pages}
  {267} (\bibinfo {year} {2012})}\BibitemShut {NoStop}%
\bibitem [{\citenamefont {Lewenstein}\ \emph {et~al.}(2012)\citenamefont
  {Lewenstein}, \citenamefont {Sanpera},\ and\ \citenamefont
  {Ahufinger}}]{Lewenstein2012}%
  \BibitemOpen
  \bibfield  {author} {\bibinfo {author} {\bibfnamefont {M.}~\bibnamefont
  {Lewenstein}}, \bibinfo {author} {\bibfnamefont {A.}~\bibnamefont {Sanpera}},
  \ and\ \bibinfo {author} {\bibfnamefont {V.}~\bibnamefont {Ahufinger}},\
  }\href@noop {} {\emph {\bibinfo {title} {Ultracold Atoms in Optical Lattices:
  Simulating quantum many-body systems}}}\ (\bibinfo  {publisher} {OUP
  Oxford},\ \bibinfo {year} {2012})\BibitemShut {NoStop}%
\bibitem [{\citenamefont {Jaksch}\ and\ \citenamefont
  {Zoller}(2005)}]{Jaksch2005}%
  \BibitemOpen
  \bibfield  {author} {\bibinfo {author} {\bibfnamefont {D.}~\bibnamefont
  {Jaksch}}\ and\ \bibinfo {author} {\bibfnamefont {P.}~\bibnamefont
  {Zoller}},\ }\href {\doibase http://dx.doi.org/10.1016/j.aop.2004.09.010}
  {\bibfield  {journal} {\bibinfo  {journal} {Ann. Phys. (N. Y.)}\ }\textbf
  {\bibinfo {volume} {315}},\ \bibinfo {pages} {52 } (\bibinfo {year}
  {2005})}\BibitemShut {NoStop}%
\bibitem [{\citenamefont {Jaksch}\ \emph {et~al.}(1998)\citenamefont {Jaksch},
  \citenamefont {Bruder}, \citenamefont {Cirac}, \citenamefont {Gardiner},\
  and\ \citenamefont {Zoller}}]{Jaksch1998}%
  \BibitemOpen
  \bibfield  {author} {\bibinfo {author} {\bibfnamefont {D.}~\bibnamefont
  {Jaksch}}, \bibinfo {author} {\bibfnamefont {C.}~\bibnamefont {Bruder}},
  \bibinfo {author} {\bibfnamefont {J.~I.}\ \bibnamefont {Cirac}}, \bibinfo
  {author} {\bibfnamefont {C.~W.}\ \bibnamefont {Gardiner}}, \ and\ \bibinfo
  {author} {\bibfnamefont {P.}~\bibnamefont {Zoller}},\ }\href {\doibase
  10.1103/PhysRevLett.81.3108} {\bibfield  {journal} {\bibinfo  {journal}
  {Phys. Rev. Lett.}\ }\textbf {\bibinfo {volume} {81}},\ \bibinfo {pages}
  {3108} (\bibinfo {year} {1998})}\BibitemShut {NoStop}%
\bibitem [{\citenamefont {Polkovnikov}\ \emph {et~al.}(2011)\citenamefont
  {Polkovnikov}, \citenamefont {Sengupta}, \citenamefont {Silva},\ and\
  \citenamefont {Vengalattore}}]{Polkovnikov2011}%
  \BibitemOpen
  \bibfield  {author} {\bibinfo {author} {\bibfnamefont {A.}~\bibnamefont
  {Polkovnikov}}, \bibinfo {author} {\bibfnamefont {K.}~\bibnamefont
  {Sengupta}}, \bibinfo {author} {\bibfnamefont {A.}~\bibnamefont {Silva}}, \
  and\ \bibinfo {author} {\bibfnamefont {M.}~\bibnamefont {Vengalattore}},\
  }\href {\doibase 10.1103/RevModPhys.83.863} {\bibfield  {journal} {\bibinfo
  {journal} {Rev. Mod. Phys.}\ }\textbf {\bibinfo {volume} {83}},\ \bibinfo
  {pages} {863} (\bibinfo {year} {2011})}\BibitemShut {NoStop}%
\bibitem [{\citenamefont {Daley}\ \emph {et~al.}(2014)\citenamefont {Daley},
  \citenamefont {Rigol},\ and\ \citenamefont {Weiss}}]{Daley2014b}%
  \BibitemOpen
  \bibfield  {author} {\bibinfo {author} {\bibfnamefont {A.~J.}\ \bibnamefont
  {Daley}}, \bibinfo {author} {\bibfnamefont {M.}~\bibnamefont {Rigol}}, \ and\
  \bibinfo {author} {\bibfnamefont {D.~S.}\ \bibnamefont {Weiss}},\ }\href
  {http://stacks.iop.org/1367-2630/16/i=9/a=095006} {\bibfield  {journal}
  {\bibinfo  {journal} {New J. Phys.}\ }\textbf {\bibinfo {volume} {16}},\
  \bibinfo {pages} {095006} (\bibinfo {year} {2014})}\BibitemShut {NoStop}%
\bibitem [{\citenamefont {M{\"u}ller}\ \emph {et~al.}(2012)\citenamefont
  {M{\"u}ller}, \citenamefont {Diehl}, \citenamefont {Pupillo},\ and\
  \citenamefont {Zoller}}]{Muller2012}%
  \BibitemOpen
  \bibfield  {author} {\bibinfo {author} {\bibfnamefont {M.}~\bibnamefont
  {M{\"u}ller}}, \bibinfo {author} {\bibfnamefont {S.}~\bibnamefont {Diehl}},
  \bibinfo {author} {\bibfnamefont {G.}~\bibnamefont {Pupillo}}, \ and\
  \bibinfo {author} {\bibfnamefont {P.}~\bibnamefont {Zoller}},\ }\href
  {https://www.sciencedirect.com/science/article/pii/B9780123964823000016}
  {\bibfield  {journal} {\bibinfo  {journal} {Advances In Atomic, Molecular,
  and Optical Physics}\ }\textbf {\bibinfo {volume} {61}},\ \bibinfo {pages}
  {1} (\bibinfo {year} {2012})}\BibitemShut {NoStop}%
\bibitem [{\citenamefont {{Choi}}\ \emph {et~al.}(2018)\citenamefont {{Choi}},
  \citenamefont {{Turner}}, \citenamefont {{Pichler}}, \citenamefont {{Ho}},
  \citenamefont {{Michailidis}}, \citenamefont {{Papi{\'c}}}, \citenamefont
  {{Serbyn}}, \citenamefont {{Lukin}},\ and\ \citenamefont
  {{Abanin}}}]{Choi2018}%
  \BibitemOpen
  \bibfield  {author} {\bibinfo {author} {\bibfnamefont {S.}~\bibnamefont
  {{Choi}}}, \bibinfo {author} {\bibfnamefont {C.~J.}\ \bibnamefont
  {{Turner}}}, \bibinfo {author} {\bibfnamefont {H.}~\bibnamefont {{Pichler}}},
  \bibinfo {author} {\bibfnamefont {W.~W.}\ \bibnamefont {{Ho}}}, \bibinfo
  {author} {\bibfnamefont {A.~A.}\ \bibnamefont {{Michailidis}}}, \bibinfo
  {author} {\bibfnamefont {Z.}~\bibnamefont {{Papi{\'c}}}}, \bibinfo {author}
  {\bibfnamefont {M.}~\bibnamefont {{Serbyn}}}, \bibinfo {author}
  {\bibfnamefont {M.~D.}\ \bibnamefont {{Lukin}}}, \ and\ \bibinfo {author}
  {\bibfnamefont {D.~A.}\ \bibnamefont {{Abanin}}},\ }\href@noop {} {\
  (\bibinfo {year} {2018})},\ \Eprint {http://arxiv.org/abs/1812.05561}
  {arXiv:1812.05561} \BibitemShut {NoStop}%
\bibitem [{\citenamefont {Ho}\ \emph {et~al.}(2019)\citenamefont {Ho},
  \citenamefont {Choi}, \citenamefont {Pichler},\ and\ \citenamefont
  {Lukin}}]{Ho2019}%
  \BibitemOpen
  \bibfield  {author} {\bibinfo {author} {\bibfnamefont {W.~W.}\ \bibnamefont
  {Ho}}, \bibinfo {author} {\bibfnamefont {S.}~\bibnamefont {Choi}}, \bibinfo
  {author} {\bibfnamefont {H.}~\bibnamefont {Pichler}}, \ and\ \bibinfo
  {author} {\bibfnamefont {M.~D.}\ \bibnamefont {Lukin}},\ }\href {\doibase
  10.1103/PhysRevLett.122.040603} {\bibfield  {journal} {\bibinfo  {journal}
  {Phys. Rev. Lett.}\ }\textbf {\bibinfo {volume} {122}},\ \bibinfo {pages}
  {040603} (\bibinfo {year} {2019})}\BibitemShut {NoStop}%
\bibitem [{\citenamefont {Samajdar}\ \emph {et~al.}(2018)\citenamefont
  {Samajdar}, \citenamefont {Choi}, \citenamefont {Pichler}, \citenamefont
  {Lukin},\ and\ \citenamefont {Sachdev}}]{Samajdar2018}%
  \BibitemOpen
  \bibfield  {author} {\bibinfo {author} {\bibfnamefont {R.}~\bibnamefont
  {Samajdar}}, \bibinfo {author} {\bibfnamefont {S.}~\bibnamefont {Choi}},
  \bibinfo {author} {\bibfnamefont {H.}~\bibnamefont {Pichler}}, \bibinfo
  {author} {\bibfnamefont {M.~D.}\ \bibnamefont {Lukin}}, \ and\ \bibinfo
  {author} {\bibfnamefont {S.}~\bibnamefont {Sachdev}},\ }\href {\doibase
  10.1103/PhysRevA.98.023614} {\bibfield  {journal} {\bibinfo  {journal} {Phys.
  Rev. A}\ }\textbf {\bibinfo {volume} {98}},\ \bibinfo {pages} {023614}
  (\bibinfo {year} {2018})}\BibitemShut {NoStop}%
\bibitem [{\citenamefont {Verresen}\ \emph {et~al.}()\citenamefont {Verresen},
  \citenamefont {Vishwanath},\ and\ \citenamefont {Pollmann}}]{Verresen2019}%
  \BibitemOpen
  \bibfield  {author} {\bibinfo {author} {\bibfnamefont {R.}~\bibnamefont
  {Verresen}}, \bibinfo {author} {\bibfnamefont {A.}~\bibnamefont
  {Vishwanath}}, \ and\ \bibinfo {author} {\bibfnamefont {F.}~\bibnamefont
  {Pollmann}},\ }\href@noop {} {\ }\Eprint {http://arxiv.org/abs/1903.09179}
  {arXiv:1903.09179} \BibitemShut {NoStop}%
\bibitem [{\citenamefont {Surace}\ \emph {et~al.}()\citenamefont {Surace},
  \citenamefont {Mazza}, \citenamefont {Giudici}, \citenamefont {Lerose},
  \citenamefont {Gambassi},\ and\ \citenamefont {Dalmonte}}]{Surace2019}%
  \BibitemOpen
  \bibfield  {author} {\bibinfo {author} {\bibfnamefont {F.~M.}\ \bibnamefont
  {Surace}}, \bibinfo {author} {\bibfnamefont {P.~P.}\ \bibnamefont {Mazza}},
  \bibinfo {author} {\bibfnamefont {G.}~\bibnamefont {Giudici}}, \bibinfo
  {author} {\bibfnamefont {A.}~\bibnamefont {Lerose}}, \bibinfo {author}
  {\bibfnamefont {A.}~\bibnamefont {Gambassi}}, \ and\ \bibinfo {author}
  {\bibfnamefont {M.}~\bibnamefont {Dalmonte}},\ }\href@noop {} {\ }\Eprint
  {http://arxiv.org/abs/1902.09551} {arXiv:1902.09551} \BibitemShut {NoStop}%
\bibitem [{\citenamefont {Greiner}\ \emph {et~al.}(2002)\citenamefont
  {Greiner}, \citenamefont {Mandel}, \citenamefont {Esslinger}, \citenamefont
  {H\"ansch},\ and\ \citenamefont {Bloch}}]{Greiner2002}%
  \BibitemOpen
  \bibfield  {author} {\bibinfo {author} {\bibfnamefont {M.}~\bibnamefont
  {Greiner}}, \bibinfo {author} {\bibfnamefont {O.}~\bibnamefont {Mandel}},
  \bibinfo {author} {\bibfnamefont {T.}~\bibnamefont {Esslinger}}, \bibinfo
  {author} {\bibfnamefont {T.~W.}\ \bibnamefont {H\"ansch}}, \ and\ \bibinfo
  {author} {\bibfnamefont {I.}~\bibnamefont {Bloch}},\ }\href {\doibase
  10.1038/415039a} {\bibfield  {journal} {\bibinfo  {journal} {Nature}\
  }\textbf {\bibinfo {volume} {415}},\ \bibinfo {pages} {39} (\bibinfo {year}
  {2002})}\BibitemShut {NoStop}%
\bibitem [{\citenamefont {Simon}\ \emph {et~al.}(2011)\citenamefont {Simon},
  \citenamefont {Bakr}, \citenamefont {Ma}, \citenamefont {Tai}, \citenamefont
  {Preiss},\ and\ \citenamefont {Greiner}}]{Simon2011}%
  \BibitemOpen
  \bibfield  {author} {\bibinfo {author} {\bibfnamefont {J.}~\bibnamefont
  {Simon}}, \bibinfo {author} {\bibfnamefont {W.~S.}\ \bibnamefont {Bakr}},
  \bibinfo {author} {\bibfnamefont {R.}~\bibnamefont {Ma}}, \bibinfo {author}
  {\bibfnamefont {M.~E.}\ \bibnamefont {Tai}}, \bibinfo {author} {\bibfnamefont
  {P.~M.}\ \bibnamefont {Preiss}}, \ and\ \bibinfo {author} {\bibfnamefont
  {M.}~\bibnamefont {Greiner}},\ }\href {\doibase 10.1038/nature09994}
  {\bibfield  {journal} {\bibinfo  {journal} {Nature}\ }\textbf {\bibinfo
  {volume} {472}},\ \bibinfo {pages} {307} (\bibinfo {year}
  {2011})}\BibitemShut {NoStop}%
\bibitem [{\citenamefont {Meinert}\ \emph {et~al.}(2013)\citenamefont
  {Meinert}, \citenamefont {Mark}, \citenamefont {Kirilov}, \citenamefont
  {Lauber}, \citenamefont {Weinmann}, \citenamefont {Daley},\ and\
  \citenamefont {N\"agerl}}]{Meinert2013}%
  \BibitemOpen
  \bibfield  {author} {\bibinfo {author} {\bibfnamefont {F.}~\bibnamefont
  {Meinert}}, \bibinfo {author} {\bibfnamefont {M.~J.}\ \bibnamefont {Mark}},
  \bibinfo {author} {\bibfnamefont {E.}~\bibnamefont {Kirilov}}, \bibinfo
  {author} {\bibfnamefont {K.}~\bibnamefont {Lauber}}, \bibinfo {author}
  {\bibfnamefont {P.}~\bibnamefont {Weinmann}}, \bibinfo {author}
  {\bibfnamefont {A.~J.}\ \bibnamefont {Daley}}, \ and\ \bibinfo {author}
  {\bibfnamefont {H.-C.}\ \bibnamefont {N\"agerl}},\ }\href {\doibase
  10.1103/PhysRevLett.111.053003} {\bibfield  {journal} {\bibinfo  {journal}
  {Phys. Rev. Lett.}\ }\textbf {\bibinfo {volume} {111}},\ \bibinfo {pages}
  {053003} (\bibinfo {year} {2013})}\BibitemShut {NoStop}%
\bibitem [{\citenamefont {Meinert}\ \emph {et~al.}(2014)\citenamefont
  {Meinert}, \citenamefont {Mark}, \citenamefont {Kirilov}, \citenamefont
  {Lauber}, \citenamefont {Weinmann}, \citenamefont {Gr{\"o}bner},
  \citenamefont {Daley},\ and\ \citenamefont {N{\"a}gerl}}]{Meinert2014}%
  \BibitemOpen
  \bibfield  {author} {\bibinfo {author} {\bibfnamefont {F.}~\bibnamefont
  {Meinert}}, \bibinfo {author} {\bibfnamefont {M.~J.}\ \bibnamefont {Mark}},
  \bibinfo {author} {\bibfnamefont {E.}~\bibnamefont {Kirilov}}, \bibinfo
  {author} {\bibfnamefont {K.}~\bibnamefont {Lauber}}, \bibinfo {author}
  {\bibfnamefont {P.}~\bibnamefont {Weinmann}}, \bibinfo {author}
  {\bibfnamefont {M.}~\bibnamefont {Gr{\"o}bner}}, \bibinfo {author}
  {\bibfnamefont {A.~J.}\ \bibnamefont {Daley}}, \ and\ \bibinfo {author}
  {\bibfnamefont {H.-C.}\ \bibnamefont {N{\"a}gerl}},\ }\href {\doibase
  10.1126/science.1248402} {\bibfield  {journal} {\bibinfo  {journal}
  {Science}\ }\textbf {\bibinfo {volume} {344}},\ \bibinfo {pages} {1259}
  (\bibinfo {year} {2014})}\BibitemShut {NoStop}%
\bibitem [{\citenamefont {Sachdev}\ \emph {et~al.}(2002)\citenamefont
  {Sachdev}, \citenamefont {Sengupta},\ and\ \citenamefont
  {Girvin}}]{Sachdev2002}%
  \BibitemOpen
  \bibfield  {author} {\bibinfo {author} {\bibfnamefont {S.}~\bibnamefont
  {Sachdev}}, \bibinfo {author} {\bibfnamefont {K.}~\bibnamefont {Sengupta}}, \
  and\ \bibinfo {author} {\bibfnamefont {S.~M.}\ \bibnamefont {Girvin}},\
  }\href {\doibase 10.1103/PhysRevB.66.075128} {\bibfield  {journal} {\bibinfo
  {journal} {Phys. Rev. B}\ }\textbf {\bibinfo {volume} {66}},\ \bibinfo
  {pages} {075128} (\bibinfo {year} {2002})}\BibitemShut {NoStop}%
\bibitem [{\citenamefont {Buyskikh}\ \emph {et~al.}(2019)\citenamefont
  {Buyskikh}, \citenamefont {Tagliacozzo}, \citenamefont {Schuricht},
  \citenamefont {Hooley}, \citenamefont {Pekker},\ and\ \citenamefont
  {Daley}}]{Buyskikh2018a}%
  \BibitemOpen
  \bibfield  {author} {\bibinfo {author} {\bibfnamefont {A.~S.}\ \bibnamefont
  {Buyskikh}}, \bibinfo {author} {\bibfnamefont {L.}~\bibnamefont
  {Tagliacozzo}}, \bibinfo {author} {\bibfnamefont {D.}~\bibnamefont
  {Schuricht}}, \bibinfo {author} {\bibfnamefont {C.~A.}\ \bibnamefont
  {Hooley}}, \bibinfo {author} {\bibfnamefont {D.}~\bibnamefont {Pekker}}, \
  and\ \bibinfo {author} {\bibfnamefont {A.~J.}\ \bibnamefont {Daley}},\ }\href
  {\doibase 10.1103/PhysRevA.100.023627} {\bibfield  {journal} {\bibinfo
  {journal} {Phys. Rev. A}\ }\textbf {\bibinfo {volume} {100}},\ \bibinfo
  {pages} {023627} (\bibinfo {year} {2019})}\BibitemShut {NoStop}%
\bibitem [{\citenamefont {Kolodrubetz}\ \emph {et~al.}(2012)\citenamefont
  {Kolodrubetz}, \citenamefont {Pekker}, \citenamefont {Clark},\ and\
  \citenamefont {Sengupta}}]{Kolodrubetz2012}%
  \BibitemOpen
  \bibfield  {author} {\bibinfo {author} {\bibfnamefont {M.}~\bibnamefont
  {Kolodrubetz}}, \bibinfo {author} {\bibfnamefont {D.}~\bibnamefont {Pekker}},
  \bibinfo {author} {\bibfnamefont {B.~K.}\ \bibnamefont {Clark}}, \ and\
  \bibinfo {author} {\bibfnamefont {K.}~\bibnamefont {Sengupta}},\ }\href
  {\doibase 10.1103/PhysRevB.85.100505} {\bibfield  {journal} {\bibinfo
  {journal} {Phys. Rev. B}\ }\textbf {\bibinfo {volume} {85}},\ \bibinfo
  {pages} {100505} (\bibinfo {year} {2012})}\BibitemShut {NoStop}%
\bibitem [{\citenamefont {Mu\~noz Arias}\ \emph {et~al.}(2016)\citenamefont
  {Mu\~noz Arias}, \citenamefont {Madro\~nero},\ and\ \citenamefont
  {Parra-Murillo}}]{Munoz-Arias2016}%
  \BibitemOpen
  \bibfield  {author} {\bibinfo {author} {\bibfnamefont {M.~H.}\ \bibnamefont
  {Mu\~noz Arias}}, \bibinfo {author} {\bibfnamefont {J.}~\bibnamefont
  {Madro\~nero}}, \ and\ \bibinfo {author} {\bibfnamefont {C.~A.}\ \bibnamefont
  {Parra-Murillo}},\ }\href {\doibase 10.1103/PhysRevA.93.043603} {\bibfield
  {journal} {\bibinfo  {journal} {Phys. Rev. A}\ }\textbf {\bibinfo {volume}
  {93}},\ \bibinfo {pages} {043603} (\bibinfo {year} {2016})}\BibitemShut
  {NoStop}%
\bibitem [{\citenamefont {Ovchinnikov}\ \emph {et~al.}(2003)\citenamefont
  {Ovchinnikov}, \citenamefont {Dmitriev}, \citenamefont {Krivnov},\ and\
  \citenamefont {Cheranovskii}}]{Ovchinnikov2003}%
  \BibitemOpen
  \bibfield  {author} {\bibinfo {author} {\bibfnamefont {A.~A.}\ \bibnamefont
  {Ovchinnikov}}, \bibinfo {author} {\bibfnamefont {D.~V.}\ \bibnamefont
  {Dmitriev}}, \bibinfo {author} {\bibfnamefont {V.~Y.}\ \bibnamefont
  {Krivnov}}, \ and\ \bibinfo {author} {\bibfnamefont {V.~O.}\ \bibnamefont
  {Cheranovskii}},\ }\href {\doibase 10.1103/PhysRevB.68.214406} {\bibfield
  {journal} {\bibinfo  {journal} {Phys. Rev. B}\ }\textbf {\bibinfo {volume}
  {68}},\ \bibinfo {pages} {214406} (\bibinfo {year} {2003})}\BibitemShut
  {NoStop}%
\bibitem [{\citenamefont {Schollwoeck}(2011)}]{Schollwoeck2011}%
  \BibitemOpen
  \bibfield  {author} {\bibinfo {author} {\bibfnamefont {U.}~\bibnamefont
  {Schollwoeck}},\ }\href {\doibase 10.1016/j.aop.2010.09.012} {\bibfield
  {journal} {\bibinfo  {journal} {Ann. Phys. (N. Y.)}\ }\textbf {\bibinfo
  {volume} {326}},\ \bibinfo {pages} {96} (\bibinfo {year} {2011})}\BibitemShut
  {NoStop}%
\bibitem [{\citenamefont {Haegeman}\ \emph {et~al.}(2011)\citenamefont
  {Haegeman}, \citenamefont {Cirac}, \citenamefont {Osborne}, \citenamefont
  {Pi{\v z}orn}, \citenamefont {Verschelde},\ and\ \citenamefont
  {Verstraete}}]{Haegeman2011}%
  \BibitemOpen
  \bibfield  {author} {\bibinfo {author} {\bibfnamefont {J.}~\bibnamefont
  {Haegeman}}, \bibinfo {author} {\bibfnamefont {J.~I.}\ \bibnamefont {Cirac}},
  \bibinfo {author} {\bibfnamefont {T.~J.}\ \bibnamefont {Osborne}}, \bibinfo
  {author} {\bibfnamefont {I.}~\bibnamefont {Pi{\v z}orn}}, \bibinfo {author}
  {\bibfnamefont {H.}~\bibnamefont {Verschelde}}, \ and\ \bibinfo {author}
  {\bibfnamefont {F.}~\bibnamefont {Verstraete}},\ }\href {\doibase
  10.1103/PhysRevLett.107.070601} {\bibfield  {journal} {\bibinfo  {journal}
  {Phys. Rev. Lett.}\ }\textbf {\bibinfo {volume} {107}},\ \bibinfo {pages}
  {070601} (\bibinfo {year} {2011})}\BibitemShut {NoStop}%
\bibitem [{\citenamefont {Koffel}\ \emph {et~al.}(2012)\citenamefont {Koffel},
  \citenamefont {Lewenstein},\ and\ \citenamefont {Tagliacozzo}}]{Koffel2012}%
  \BibitemOpen
  \bibfield  {author} {\bibinfo {author} {\bibfnamefont {T.}~\bibnamefont
  {Koffel}}, \bibinfo {author} {\bibfnamefont {M.}~\bibnamefont {Lewenstein}},
  \ and\ \bibinfo {author} {\bibfnamefont {L.}~\bibnamefont {Tagliacozzo}},\
  }\href {\doibase 10.1103/PhysRevLett.109.267203} {\bibfield  {journal}
  {\bibinfo  {journal} {Phys. Rev. Lett.}\ }\textbf {\bibinfo {volume} {109}},\
  \bibinfo {pages} {267203} (\bibinfo {year} {2012})}\BibitemShut {NoStop}%
\bibitem [{\citenamefont {Haegeman}\ \emph {et~al.}(2013)\citenamefont
  {Haegeman}, \citenamefont {Osborne},\ and\ \citenamefont
  {Verstraete}}]{Haegeman2013}%
  \BibitemOpen
  \bibfield  {author} {\bibinfo {author} {\bibfnamefont {J.}~\bibnamefont
  {Haegeman}}, \bibinfo {author} {\bibfnamefont {T.~J.}\ \bibnamefont
  {Osborne}}, \ and\ \bibinfo {author} {\bibfnamefont {F.}~\bibnamefont
  {Verstraete}},\ }\href {\doibase 10.1103/PhysRevB.88.075133} {\bibfield
  {journal} {\bibinfo  {journal} {Phys. Rev. B}\ }\textbf {\bibinfo {volume}
  {88}},\ \bibinfo {pages} {075133} (\bibinfo {year} {2013})}\BibitemShut
  {NoStop}%
\bibitem [{\citenamefont {Haegeman}\ \emph {et~al.}(2016)\citenamefont
  {Haegeman}, \citenamefont {Lubich}, \citenamefont {Oseledets}, \citenamefont
  {Vandereycken},\ and\ \citenamefont {Verstraete}}]{Haegeman2016}%
  \BibitemOpen
  \bibfield  {author} {\bibinfo {author} {\bibfnamefont {J.}~\bibnamefont
  {Haegeman}}, \bibinfo {author} {\bibfnamefont {C.}~\bibnamefont {Lubich}},
  \bibinfo {author} {\bibfnamefont {I.}~\bibnamefont {Oseledets}}, \bibinfo
  {author} {\bibfnamefont {B.}~\bibnamefont {Vandereycken}}, \ and\ \bibinfo
  {author} {\bibfnamefont {F.}~\bibnamefont {Verstraete}},\ }\href {\doibase
  10.1103/PhysRevB.94.165116} {\bibfield  {journal} {\bibinfo  {journal} {Phys.
  Rev. B}\ }\textbf {\bibinfo {volume} {94}},\ \bibinfo {pages} {165116}
  (\bibinfo {year} {2016})}\BibitemShut {NoStop}%
\bibitem [{\citenamefont {Verstraete}\ \emph {et~al.}(2004)\citenamefont
  {Verstraete}, \citenamefont {Garc\'{\i}a-Ripoll},\ and\ \citenamefont
  {Cirac}}]{Verstraete2004}%
  \BibitemOpen
  \bibfield  {author} {\bibinfo {author} {\bibfnamefont {F.}~\bibnamefont
  {Verstraete}}, \bibinfo {author} {\bibfnamefont {J.~J.}\ \bibnamefont
  {Garc\'{\i}a-Ripoll}}, \ and\ \bibinfo {author} {\bibfnamefont {J.~I.}\
  \bibnamefont {Cirac}},\ }\href {\doibase 10.1103/PhysRevLett.93.207204}
  {\bibfield  {journal} {\bibinfo  {journal} {Phys. Rev. Lett.}\ }\textbf
  {\bibinfo {volume} {93}},\ \bibinfo {pages} {207204} (\bibinfo {year}
  {2004})}\BibitemShut {NoStop}%
\bibitem [{\citenamefont {las Cuevas}\ \emph {et~al.}(2013)\citenamefont {las
  Cuevas}, \citenamefont {Schuch}, \citenamefont {P{\'{e}}rez-Garc{\'{\i}}a},\
  and\ \citenamefont {Cirac}}]{Cuevas2013}%
  \BibitemOpen
  \bibfield  {author} {\bibinfo {author} {\bibfnamefont {G.~D.}\ \bibnamefont
  {las Cuevas}}, \bibinfo {author} {\bibfnamefont {N.}~\bibnamefont {Schuch}},
  \bibinfo {author} {\bibfnamefont {D.}~\bibnamefont
  {P{\'{e}}rez-Garc{\'{\i}}a}}, \ and\ \bibinfo {author} {\bibfnamefont
  {J.~I.}\ \bibnamefont {Cirac}},\ }\href {\doibase
  10.1088/1367-2630/15/12/123021} {\bibfield  {journal} {\bibinfo  {journal}
  {New J. Phys.}\ }\textbf {\bibinfo {volume} {15}},\ \bibinfo {pages} {123021}
  (\bibinfo {year} {2013})}\BibitemShut {NoStop}%
\bibitem [{\citenamefont {Binder}(1981)}]{Binder1981}%
  \BibitemOpen
  \bibfield  {author} {\bibinfo {author} {\bibfnamefont {K.}~\bibnamefont
  {Binder}},\ }\href {\doibase 10.1007/BF01293604} {\bibfield  {journal}
  {\bibinfo  {journal} {Z. Physik B - Condensed Matter}\ }\textbf {\bibinfo
  {volume} {43}},\ \bibinfo {pages} {119} (\bibinfo {year} {1981})}\BibitemShut
  {NoStop}%
\bibitem [{\citenamefont {Callan}\ and\ \citenamefont
  {Wilczek}(1994)}]{Callan1994}%
  \BibitemOpen
  \bibfield  {author} {\bibinfo {author} {\bibfnamefont {C.}~\bibnamefont
  {Callan}}\ and\ \bibinfo {author} {\bibfnamefont {F.}~\bibnamefont
  {Wilczek}},\ }\href {\doibase https://doi.org/10.1016/0370-2693(94)91007-3}
  {\bibfield  {journal} {\bibinfo  {journal} {Phys. Lett. B}\ }\textbf
  {\bibinfo {volume} {333}},\ \bibinfo {pages} {55 } (\bibinfo {year}
  {1994})}\BibitemShut {NoStop}%
\bibitem [{\citenamefont {Osborne}\ and\ \citenamefont
  {Nielsen}(2002)}]{Osborne2002}%
  \BibitemOpen
  \bibfield  {author} {\bibinfo {author} {\bibfnamefont {T.~J.}\ \bibnamefont
  {Osborne}}\ and\ \bibinfo {author} {\bibfnamefont {M.~A.}\ \bibnamefont
  {Nielsen}},\ }\href {\doibase 10.1103/PhysRevA.66.032110} {\bibfield
  {journal} {\bibinfo  {journal} {Phys. Rev. A}\ }\textbf {\bibinfo {volume}
  {66}},\ \bibinfo {pages} {032110} (\bibinfo {year} {2002})}\BibitemShut
  {NoStop}%
\bibitem [{\citenamefont {Vidal}\ \emph {et~al.}(2003)\citenamefont {Vidal},
  \citenamefont {Latorre}, \citenamefont {Rico},\ and\ \citenamefont
  {Kitaev}}]{Vidal2003}%
  \BibitemOpen
  \bibfield  {author} {\bibinfo {author} {\bibfnamefont {G.}~\bibnamefont
  {Vidal}}, \bibinfo {author} {\bibfnamefont {J.~I.}\ \bibnamefont {Latorre}},
  \bibinfo {author} {\bibfnamefont {E.}~\bibnamefont {Rico}}, \ and\ \bibinfo
  {author} {\bibfnamefont {A.}~\bibnamefont {Kitaev}},\ }\href {\doibase
  10.1103/PhysRevLett.90.227902} {\bibfield  {journal} {\bibinfo  {journal}
  {Phys. Rev. Lett.}\ }\textbf {\bibinfo {volume} {90}},\ \bibinfo {pages}
  {227902} (\bibinfo {year} {2003})}\BibitemShut {NoStop}%
\bibitem [{\citenamefont {Calabrese}\ and\ \citenamefont
  {Cardy}(2004)}]{Calabrese2004}%
  \BibitemOpen
  \bibfield  {author} {\bibinfo {author} {\bibfnamefont {P.}~\bibnamefont
  {Calabrese}}\ and\ \bibinfo {author} {\bibfnamefont {J.}~\bibnamefont
  {Cardy}},\ }\href {\doibase 10.1088/1742-5468/2004/06/p06002} {\bibfield
  {journal} {\bibinfo  {journal} {J. Stat. Mech. Theory Exp.}\ }\textbf
  {\bibinfo {volume} {2004}},\ \bibinfo {pages} {P06002} (\bibinfo {year}
  {2004})}\BibitemShut {NoStop}%
\bibitem [{\citenamefont {Affleck}(1986)}]{Affleck1986}%
  \BibitemOpen
  \bibfield  {author} {\bibinfo {author} {\bibfnamefont {I.}~\bibnamefont
  {Affleck}},\ }\href {\doibase 10.1103/PhysRevLett.56.746} {\bibfield
  {journal} {\bibinfo  {journal} {Phys. Rev. Lett.}\ }\textbf {\bibinfo
  {volume} {56}},\ \bibinfo {pages} {746} (\bibinfo {year} {1986})}\BibitemShut
  {NoStop}%
\bibitem [{\citenamefont {Bl\"ote}\ \emph {et~al.}(1986)\citenamefont
  {Bl\"ote}, \citenamefont {Cardy},\ and\ \citenamefont
  {Nightingale}}]{Blote1986}%
  \BibitemOpen
  \bibfield  {author} {\bibinfo {author} {\bibfnamefont {H.~W.~J.}\
  \bibnamefont {Bl\"ote}}, \bibinfo {author} {\bibfnamefont {J.~L.}\
  \bibnamefont {Cardy}}, \ and\ \bibinfo {author} {\bibfnamefont {M.~P.}\
  \bibnamefont {Nightingale}},\ }\href {\doibase 10.1103/PhysRevLett.56.742}
  {\bibfield  {journal} {\bibinfo  {journal} {Phys. Rev. Lett.}\ }\textbf
  {\bibinfo {volume} {56}},\ \bibinfo {pages} {742} (\bibinfo {year}
  {1986})}\BibitemShut {NoStop}%
\bibitem [{\citenamefont {Cardy}(2010)}]{Cardy2010a}%
  \BibitemOpen
  \bibfield  {author} {\bibinfo {author} {\bibfnamefont {J.}~\bibnamefont
  {Cardy}},\ }\href {\doibase 10.1088/1742-5468/2010/10/P10004} {\bibfield
  {journal} {\bibinfo  {journal} {J. Stat. Mech.}\ }\textbf {\bibinfo {volume}
  {2010}},\ \bibinfo {pages} {P10004} (\bibinfo {year} {2010})}\BibitemShut
  {NoStop}%
\bibitem [{\citenamefont {Laflorencie}\ \emph {et~al.}(2006)\citenamefont
  {Laflorencie}, \citenamefont {S{\o}rensen}, \citenamefont {Chang},\ and\
  \citenamefont {Affleck}}]{Laflorencie2006}%
  \BibitemOpen
  \bibfield  {author} {\bibinfo {author} {\bibfnamefont {N.}~\bibnamefont
  {Laflorencie}}, \bibinfo {author} {\bibfnamefont {E.~S.}\ \bibnamefont
  {S{\o}rensen}}, \bibinfo {author} {\bibfnamefont {M.-S.}\ \bibnamefont
  {Chang}}, \ and\ \bibinfo {author} {\bibfnamefont {I.}~\bibnamefont
  {Affleck}},\ }\href {\doibase 10.1103/PhysRevLett.96.100603} {\bibfield
  {journal} {\bibinfo  {journal} {Phys. Rev. Lett.}\ }\textbf {\bibinfo
  {volume} {96}},\ \bibinfo {pages} {100603} (\bibinfo {year}
  {2006})}\BibitemShut {NoStop}%
\bibitem [{\citenamefont {Calabrese}\ \emph {et~al.}(2010)\citenamefont
  {Calabrese}, \citenamefont {Campostrini}, \citenamefont {Essler},\ and\
  \citenamefont {Nienhuis}}]{Calabrese2010}%
  \BibitemOpen
  \bibfield  {author} {\bibinfo {author} {\bibfnamefont {P.}~\bibnamefont
  {Calabrese}}, \bibinfo {author} {\bibfnamefont {M.}~\bibnamefont
  {Campostrini}}, \bibinfo {author} {\bibfnamefont {F.}~\bibnamefont {Essler}},
  \ and\ \bibinfo {author} {\bibfnamefont {B.}~\bibnamefont {Nienhuis}},\
  }\href {\doibase 10.1103/PhysRevLett.104.095701} {\bibfield  {journal}
  {\bibinfo  {journal} {Phys. Rev. Lett.}\ }\textbf {\bibinfo {volume} {104}},\
  \bibinfo {pages} {095701} (\bibinfo {year} {2010})}\BibitemShut {NoStop}%
\bibitem [{\citenamefont {Calabrese}\ and\ \citenamefont
  {Essler}(2010)}]{Calabrese2010a}%
  \BibitemOpen
  \bibfield  {author} {\bibinfo {author} {\bibfnamefont {P.}~\bibnamefont
  {Calabrese}}\ and\ \bibinfo {author} {\bibfnamefont {F.~H.~L.}\ \bibnamefont
  {Essler}},\ }\href {\doibase 10.1088/1742-5468/2010/08/P08029} {\bibfield
  {journal} {\bibinfo  {journal} {J. Stat. Mech.}\ }\textbf {\bibinfo {volume}
  {2010}},\ \bibinfo {pages} {P08029} (\bibinfo {year} {2010})}\BibitemShut
  {NoStop}%
\bibitem [{\citenamefont {Cardy}\ and\ \citenamefont
  {Calabrese}(2010)}]{Cardy2010}%
  \BibitemOpen
  \bibfield  {author} {\bibinfo {author} {\bibfnamefont {J.}~\bibnamefont
  {Cardy}}\ and\ \bibinfo {author} {\bibfnamefont {P.}~\bibnamefont
  {Calabrese}},\ }\href {\doibase 10.1088/1742-5468/2010/04/P04023} {\bibfield
  {journal} {\bibinfo  {journal} {J. Stat. Mech. Theory Exp.}\ }\textbf
  {\bibinfo {volume} {2010}},\ \bibinfo {pages} {P04023} (\bibinfo {year}
  {2010})}\BibitemShut {NoStop}%
\bibitem [{\citenamefont {Xavier}\ and\ \citenamefont
  {Alcaraz}(2012)}]{Xavier2012}%
  \BibitemOpen
  \bibfield  {author} {\bibinfo {author} {\bibfnamefont {J.~C.}\ \bibnamefont
  {Xavier}}\ and\ \bibinfo {author} {\bibfnamefont {F.~C.}\ \bibnamefont
  {Alcaraz}},\ }\href {\doibase 10.1103/PhysRevB.85.024418} {\bibfield
  {journal} {\bibinfo  {journal} {Phys. Rev. B}\ }\textbf {\bibinfo {volume}
  {85}},\ \bibinfo {pages} {024418} (\bibinfo {year} {2012})}\BibitemShut
  {NoStop}%
\bibitem [{\citenamefont {Wall}\ and\ \citenamefont {Carr}(2012)}]{Wall2012}%
  \BibitemOpen
  \bibfield  {author} {\bibinfo {author} {\bibfnamefont {M.~L.}\ \bibnamefont
  {Wall}}\ and\ \bibinfo {author} {\bibfnamefont {L.~D.}\ \bibnamefont
  {Carr}},\ }\href {http://stacks.iop.org/1367-2630/14/i=12/a=125015}
  {\bibfield  {journal} {\bibinfo  {journal} {New J. Phys.}\ }\textbf {\bibinfo
  {volume} {14}},\ \bibinfo {pages} {125015} (\bibinfo {year}
  {2012})}\BibitemShut {NoStop}%
\bibitem [{Note1()}]{Note1}%
  \BibitemOpen
  \bibinfo {note} {This method was initially proposed in the supplementary
  material of \cite {Koffel2012}.}\BibitemShut {Stop}%
\bibitem [{\citenamefont {Belavin}\ \emph {et~al.}(1984)\citenamefont
  {Belavin}, \citenamefont {Polyakov},\ and\ \citenamefont
  {Zamolodchikov}}]{Belavin1984}%
  \BibitemOpen
  \bibfield  {author} {\bibinfo {author} {\bibfnamefont {A.}~\bibnamefont
  {Belavin}}, \bibinfo {author} {\bibfnamefont {A.}~\bibnamefont {Polyakov}}, \
  and\ \bibinfo {author} {\bibfnamefont {A.}~\bibnamefont {Zamolodchikov}},\
  }\href {\doibase http://dx.doi.org/10.1016/0550-3213(84)90052-X} {\bibfield
  {journal} {\bibinfo  {journal} {Nucl. Phys. B}\ }\textbf {\bibinfo {volume}
  {241}},\ \bibinfo {pages} {333 } (\bibinfo {year} {1984})}\BibitemShut
  {NoStop}%
\bibitem [{\citenamefont {Henkel}(1999)}]{Henkel1999}%
  \BibitemOpen
  \bibfield  {author} {\bibinfo {author} {\bibfnamefont {M.}~\bibnamefont
  {Henkel}},\ }\href@noop {} {\emph {\bibinfo {title} {Conformal Invariance and
  Critical Phenomena}}},\ Theoretical and Mathematical Physics\ (\bibinfo
  {publisher} {Springer-Verlag},\ \bibinfo {address} {Berlin Heidelberg},\
  \bibinfo {year} {1999})\BibitemShut {NoStop}%
\bibitem [{\citenamefont {Francesco}\ \emph {et~al.}(1997)\citenamefont
  {Francesco}, \citenamefont {Mathieu},\ and\ \citenamefont
  {S{\'e}n{\'e}chal}}]{Francesco1997}%
  \BibitemOpen
  \bibfield  {author} {\bibinfo {author} {\bibfnamefont {P.}~\bibnamefont
  {Francesco}}, \bibinfo {author} {\bibfnamefont {P.}~\bibnamefont {Mathieu}},
  \ and\ \bibinfo {author} {\bibfnamefont {D.}~\bibnamefont
  {S{\'e}n{\'e}chal}},\ }\href@noop {} {\emph {\bibinfo {title} {Conformal
  Field Theory}}},\ Graduate Texts in Contemporary Physics\ (\bibinfo
  {publisher} {Springer-Verlag},\ \bibinfo {address} {New York},\ \bibinfo
  {year} {1997})\BibitemShut {NoStop}%
\bibitem [{\citenamefont {Cardy}(2008)}]{Cardy2008}%
  \BibitemOpen
  \bibfield  {author} {\bibinfo {author} {\bibfnamefont {J.~L.}\ \bibnamefont
  {Cardy}},\ }\href {\doibase 10.1140/epjb/e2008-00102-5} {\bibfield  {journal}
  {\bibinfo  {journal} {Eur. Phys. J. B}\ }\textbf {\bibinfo {volume} {64}},\
  \bibinfo {pages} {321} (\bibinfo {year} {2008})}\BibitemShut {NoStop}%
\bibitem [{\citenamefont {Fendley}\ \emph {et~al.}(2004)\citenamefont
  {Fendley}, \citenamefont {Sengupta},\ and\ \citenamefont
  {Sachdev}}]{Fendley2004}%
  \BibitemOpen
  \bibfield  {author} {\bibinfo {author} {\bibfnamefont {P.}~\bibnamefont
  {Fendley}}, \bibinfo {author} {\bibfnamefont {K.}~\bibnamefont {Sengupta}}, \
  and\ \bibinfo {author} {\bibfnamefont {S.}~\bibnamefont {Sachdev}},\ }\href
  {\doibase 10.1103/PhysRevB.69.075106} {\bibfield  {journal} {\bibinfo
  {journal} {Phys. Rev. B}\ }\textbf {\bibinfo {volume} {69}},\ \bibinfo
  {pages} {075106} (\bibinfo {year} {2004})}\BibitemShut {NoStop}%
\bibitem [{\citenamefont {Chepiga}\ and\ \citenamefont
  {Mila}(2019)}]{Chepiga2019}%
  \BibitemOpen
  \bibfield  {author} {\bibinfo {author} {\bibfnamefont {N.}~\bibnamefont
  {Chepiga}}\ and\ \bibinfo {author} {\bibfnamefont {F.}~\bibnamefont {Mila}},\
  }\href {\doibase 10.1103/PhysRevLett.122.017205} {\bibfield  {journal}
  {\bibinfo  {journal} {Phys. Rev. Lett.}\ }\textbf {\bibinfo {volume} {122}},\
  \bibinfo {pages} {017205} (\bibinfo {year} {2019})}\BibitemShut {NoStop}%
\bibitem [{\citenamefont {De~Grandi}\ \emph {et~al.}(2011)\citenamefont
  {De~Grandi}, \citenamefont {Polkovnikov},\ and\ \citenamefont
  {Sandvik}}]{DeGrandi2011}%
  \BibitemOpen
  \bibfield  {author} {\bibinfo {author} {\bibfnamefont {C.}~\bibnamefont
  {De~Grandi}}, \bibinfo {author} {\bibfnamefont {A.}~\bibnamefont
  {Polkovnikov}}, \ and\ \bibinfo {author} {\bibfnamefont {A.~W.}\ \bibnamefont
  {Sandvik}},\ }\href {\doibase 10.1103/PhysRevB.84.224303} {\bibfield
  {journal} {\bibinfo  {journal} {Phys. Rev. B}\ }\textbf {\bibinfo {volume}
  {84}},\ \bibinfo {pages} {224303} (\bibinfo {year} {2011})}\BibitemShut
  {NoStop}%
\bibitem [{\citenamefont {Rubbo}\ \emph {et~al.}(2011)\citenamefont {Rubbo},
  \citenamefont {Manmana}, \citenamefont {Peden}, \citenamefont {Holland},\
  and\ \citenamefont {Rey}}]{Rubbo2011}%
  \BibitemOpen
  \bibfield  {author} {\bibinfo {author} {\bibfnamefont {C.~P.}\ \bibnamefont
  {Rubbo}}, \bibinfo {author} {\bibfnamefont {S.~R.}\ \bibnamefont {Manmana}},
  \bibinfo {author} {\bibfnamefont {B.~M.}\ \bibnamefont {Peden}}, \bibinfo
  {author} {\bibfnamefont {M.~J.}\ \bibnamefont {Holland}}, \ and\ \bibinfo
  {author} {\bibfnamefont {A.~M.}\ \bibnamefont {Rey}},\ }\href {\doibase
  10.1103/PhysRevA.84.033638} {\bibfield  {journal} {\bibinfo  {journal} {Phys.
  Rev. A}\ }\textbf {\bibinfo {volume} {84}},\ \bibinfo {pages} {033638}
  (\bibinfo {year} {2011})}\BibitemShut {NoStop}%
\bibitem [{\citenamefont {Pielawa}\ \emph {et~al.}(2012)\citenamefont
  {Pielawa}, \citenamefont {Berg},\ and\ \citenamefont
  {Sachdev}}]{Pielawa2012}%
  \BibitemOpen
  \bibfield  {author} {\bibinfo {author} {\bibfnamefont {S.}~\bibnamefont
  {Pielawa}}, \bibinfo {author} {\bibfnamefont {E.}~\bibnamefont {Berg}}, \
  and\ \bibinfo {author} {\bibfnamefont {S.}~\bibnamefont {Sachdev}},\ }\href
  {\doibase 10.1103/PhysRevB.86.184435} {\bibfield  {journal} {\bibinfo
  {journal} {Phys. Rev. B}\ }\textbf {\bibinfo {volume} {86}},\ \bibinfo
  {pages} {184435} (\bibinfo {year} {2012})}\BibitemShut {NoStop}%
\bibitem [{\citenamefont {Kolovsky}(2016)}]{Kolovsky2016}%
  \BibitemOpen
  \bibfield  {author} {\bibinfo {author} {\bibfnamefont {A.~R.}\ \bibnamefont
  {Kolovsky}},\ }\href {\doibase 10.1103/PhysRevA.93.033626} {\bibfield
  {journal} {\bibinfo  {journal} {Phys. Rev. A}\ }\textbf {\bibinfo {volume}
  {93}},\ \bibinfo {pages} {033626} (\bibinfo {year} {2016})}\BibitemShut
  {NoStop}%
\bibitem [{boo()}]{bootstrapcollaboration}%
  \BibitemOpen
  \href {http://bootstrapcollaboration.com} {\enquote {\bibinfo {title} {Simons
  collaboration on the nonperturbative bootstrap},}\ }\BibitemShut {NoStop}%
\bibitem [{\citenamefont {Poland}\ \emph {et~al.}(2019)\citenamefont {Poland},
  \citenamefont {Rychkov},\ and\ \citenamefont {Vichi}}]{Poland2019}%
  \BibitemOpen
  \bibfield  {author} {\bibinfo {author} {\bibfnamefont {D.}~\bibnamefont
  {Poland}}, \bibinfo {author} {\bibfnamefont {S.}~\bibnamefont {Rychkov}}, \
  and\ \bibinfo {author} {\bibfnamefont {A.}~\bibnamefont {Vichi}},\ }\href
  {\doibase 10.1103/RevModPhys.91.015002} {\bibfield  {journal} {\bibinfo
  {journal} {Rev. Mod. Phys.}\ }\textbf {\bibinfo {volume} {91}},\ \bibinfo
  {pages} {015002} (\bibinfo {year} {2019})}\BibitemShut {NoStop}%
\bibitem [{\citenamefont {Atanasov}\ \emph {et~al.}(2018)\citenamefont
  {Atanasov}, \citenamefont {Hillman},\ and\ \citenamefont
  {Poland}}]{Atanasov2018}%
  \BibitemOpen
  \bibfield  {author} {\bibinfo {author} {\bibfnamefont {A.}~\bibnamefont
  {Atanasov}}, \bibinfo {author} {\bibfnamefont {A.}~\bibnamefont {Hillman}}, \
  and\ \bibinfo {author} {\bibfnamefont {D.}~\bibnamefont {Poland}},\ }\href
  {\doibase 10.1007/JHEP11(2018)140} {\bibfield  {journal} {\bibinfo  {journal}
  {J. High Energy Phys.}\ }\textbf {\bibinfo {volume} {2018}},\ \bibinfo
  {pages} {140} (\bibinfo {year} {2018})}\BibitemShut {NoStop}%
\bibitem [{\citenamefont {O'Brien}\ and\ \citenamefont
  {Fendley}(2018)}]{OBrien2018}%
  \BibitemOpen
  \bibfield  {author} {\bibinfo {author} {\bibfnamefont {E.}~\bibnamefont
  {O'Brien}}\ and\ \bibinfo {author} {\bibfnamefont {P.}~\bibnamefont
  {Fendley}},\ }\href {\doibase 10.1103/PhysRevLett.120.206403} {\bibfield
  {journal} {\bibinfo  {journal} {Phys. Rev. Lett.}\ }\textbf {\bibinfo
  {volume} {120}},\ \bibinfo {pages} {206403} (\bibinfo {year}
  {2018})}\BibitemShut {NoStop}%
\bibitem [{\citenamefont {Zhu}\ and\ \citenamefont {Franz}(2016)}]{Zhu2016}%
  \BibitemOpen
  \bibfield  {author} {\bibinfo {author} {\bibfnamefont {X.}~\bibnamefont
  {Zhu}}\ and\ \bibinfo {author} {\bibfnamefont {M.}~\bibnamefont {Franz}},\
  }\href {\doibase 10.1103/PhysRevB.93.195118} {\bibfield  {journal} {\bibinfo
  {journal} {Phys. Rev. B}\ }\textbf {\bibinfo {volume} {93}},\ \bibinfo
  {pages} {195118} (\bibinfo {year} {2016})}\BibitemShut {NoStop}%
\bibitem [{\citenamefont {Huse}\ and\ \citenamefont {Fisher}(1982)}]{Huse1982}%
  \BibitemOpen
  \bibfield  {author} {\bibinfo {author} {\bibfnamefont {D.~A.}\ \bibnamefont
  {Huse}}\ and\ \bibinfo {author} {\bibfnamefont {M.~E.}\ \bibnamefont
  {Fisher}},\ }\href {\doibase 10.1103/PhysRevLett.49.793} {\bibfield
  {journal} {\bibinfo  {journal} {Phys. Rev. Lett.}\ }\textbf {\bibinfo
  {volume} {49}},\ \bibinfo {pages} {793} (\bibinfo {year} {1982})}\BibitemShut
  {NoStop}%
\bibitem [{\citenamefont {Tagliacozzo}\ \emph {et~al.}(2014)\citenamefont
  {Tagliacozzo}, \citenamefont {Celi},\ and\ \citenamefont
  {Lewenstein}}]{Tagliacozzo2014}%
  \BibitemOpen
  \bibfield  {author} {\bibinfo {author} {\bibfnamefont {L.}~\bibnamefont
  {Tagliacozzo}}, \bibinfo {author} {\bibfnamefont {A.}~\bibnamefont {Celi}}, \
  and\ \bibinfo {author} {\bibfnamefont {M.}~\bibnamefont {Lewenstein}},\
  }\href {\doibase 10.1103/PhysRevX.4.041024} {\bibfield  {journal} {\bibinfo
  {journal} {Phys. Rev. X}\ }\textbf {\bibinfo {volume} {4}},\ \bibinfo {pages}
  {041024} (\bibinfo {year} {2014})}\BibitemShut {NoStop}%
\bibitem [{\citenamefont {Haegeman}\ \emph {et~al.}(2015)\citenamefont
  {Haegeman}, \citenamefont {Van~Acoleyen}, \citenamefont {Schuch},
  \citenamefont {Cirac},\ and\ \citenamefont {Verstraete}}]{Haegeman2015}%
  \BibitemOpen
  \bibfield  {author} {\bibinfo {author} {\bibfnamefont {J.}~\bibnamefont
  {Haegeman}}, \bibinfo {author} {\bibfnamefont {K.}~\bibnamefont
  {Van~Acoleyen}}, \bibinfo {author} {\bibfnamefont {N.}~\bibnamefont
  {Schuch}}, \bibinfo {author} {\bibfnamefont {J.~I.}\ \bibnamefont {Cirac}}, \
  and\ \bibinfo {author} {\bibfnamefont {F.}~\bibnamefont {Verstraete}},\
  }\href {\doibase 10.1103/PhysRevX.5.011024} {\bibfield  {journal} {\bibinfo
  {journal} {Phys. Rev. X}\ }\textbf {\bibinfo {volume} {5}},\ \bibinfo {pages}
  {011024} (\bibinfo {year} {2015})}\BibitemShut {NoStop}%
\bibitem [{\citenamefont {Rico}\ \emph {et~al.}(2014)\citenamefont {Rico},
  \citenamefont {Pichler}, \citenamefont {Dalmonte}, \citenamefont {Zoller},\
  and\ \citenamefont {Montangero}}]{Rico2014}%
  \BibitemOpen
  \bibfield  {author} {\bibinfo {author} {\bibfnamefont {E.}~\bibnamefont
  {Rico}}, \bibinfo {author} {\bibfnamefont {T.}~\bibnamefont {Pichler}},
  \bibinfo {author} {\bibfnamefont {M.}~\bibnamefont {Dalmonte}}, \bibinfo
  {author} {\bibfnamefont {P.}~\bibnamefont {Zoller}}, \ and\ \bibinfo {author}
  {\bibfnamefont {S.}~\bibnamefont {Montangero}},\ }\href {\doibase
  10.1103/PhysRevLett.112.201601} {\bibfield  {journal} {\bibinfo  {journal}
  {Phys. Rev. Lett.}\ }\textbf {\bibinfo {volume} {112}},\ \bibinfo {pages}
  {201601} (\bibinfo {year} {2014})}\BibitemShut {NoStop}%
\bibitem [{dat()}]{data_short}%
  \BibitemOpen
  \href@noop {} {}\bibinfo {note}
  {\url{https://doi.org/10.15129/53946e0e-601d-40bb-bfbb-8d237224092b}}\BibitemShut
  {NoStop}%
\end{thebibliography}%

\end{document}